\documentclass[12pt]{article}

\usepackage{amsmath}
\usepackage{amssymb}
\usepackage{latexsym}

\def\beqar {\begin{eqnarray}}
\def\eeqar {\end{eqnarray}}
\def\beq {\begin{equation}}
\def\eeq {\end{equation}}

\def\no2 {{\textstyle{n\over 2}}}

\begin{document}

\rightline{} \vspace{1cm}
\begin{center}

{\Large \bf Towards $Z_2$-protected gauge--Higgs unification}

\vspace{1.5cm}

S.~Randjbar-Daemi\footnote{The Abdus Salam International Centre
for Theoretical Physics, Trieste, Italy} and V. Rubakov
\footnote{Institute for Nuclear Research of the Russian Academy of
Sciences,
Moscow, Russia}\\

\end{center}
\vskip 0.8cm \centerline{\bf Abstract}
\bigskip
\small{In theories with flux compactification in eight or higher dimensions,
the extra-dimensional components of the gauge field may be 
regarded as the Higgs field candidates. We suggest a way to protect
these components from getting large tree-level masses by imposing
a $Z_2$-symmetry acting on compact manifolds and background
fields on them. In our scheme the infinite series of heavy
KK modes naturally decouples from the light Higgs candidates, whose
number is generically larger than one. We also present toy models
with three families of leptons, illustrating that the Yukawa sector in
our scheme is fairly strongly constrained. In one of these models,
one fermion gets a tree-level mass after electroweak symmetry breaking,
while two others remain naturally massless at the tree level.
}

\vspace{3cm}


\section{Introduction and summary}

The idea that the Higgs field(s) may be identified with
extra-dimensional component(s) of gauge fields --- gauge--Higgs
unification --- is of considerable interest for many
years~\cite{GHU-1,GHU-2,GHU-2aa},
with more recent emphasis put on the gauge hierarchy problem and
stability of the electroweak scale~\cite{GHU-2a,Seif-Gia,GHU3}.
For similar number of years it is
known~\cite{Salam, Flux} that an attrative way to obtain
chiral 4-dimensional fermions is to populate compact extra dimensions
with topologically non-trivial background gauge fields; this can
be achieved via flux compactifications. These two mechanisms,
however, appear to be in potential conflict with
each other, as the background gauge fields generically induce large
(normal or tachyonic \cite{RandjbarDaemi:1983bw}) mass terms for their perturbations tangent to
extra dimensions. As a possible way out it has been noted~\cite{Seif-Gia}
that in the case of product compact manifolds, there may occur cancellations
between different contributions to the mass terms, so that some 
4-dimensional scalars ---
components of multi-dimensional gauge fields --- may be light and even
massless. Once their masses are small and tachyonic, either at the tree level or
due to radiative corrections, the extra-dimensional components
of the gauge field perturbations become indeed the Higgs field
candidates.

In this paper we elaborate on this class of theories. We give a
simple characterization of potentially light 4-dimensional scalar fields
in those
models with flux compactification  where the geometry of extra
dimensions is that of a product of two-dimensional
compact manifolds. With this characterization, the number of
potentially light scalar fields, their tree level masses and wave functions
are straightforwardly calculable. We point out that in some cases,
zero values of the tree level masses, rather than being a result of
fine-tuning, occur as a consequence of a discrete symmetry $Z_2$
between
the manifolds entering the product. In those cases the number of massless
scalars is necessarily greater than 1. Once this discrete symmetry
is slightly broken at the classical level, 
the scalars obtain small tree level
masses, half of which are automatically tachyonic, and half are
normal. In fact, the multiplicity of the Higgs field candidates and
their positive-$m^2$ partners is  a fairly generic property of the class of
models we consider, this feature being of potential phenomenological
importance. Another generic property,
which is common to models of gauge--Higgs unification in more than 5
dimensions~\cite{Seif-Gia,GHU3,Csaki1,Scrucca1}
 is that the quartic self-couplings of
the light scalars exist already at the tree level, unlike in the simplest
versions of the Hosotani/Scherk--Schwarz mechanism where self-couplings are
induced radiatively and hence are often too 
low~\cite{masses,masses2,masses3,masses4}. We will further comment
on phenomenology of our models towards the end of this paper.

We then proceed by giving concrete examples showing that
it is relatively straightforward to obtain a pattern of the Higgs
field and chiral fermion representations resembling that of the
Standard Model.
Our first example, however, illustrates one of
the  obstacles for
utilizing our construction for model building with simple gauge groups.
Namely,
in this example, no Yukawa interactions between the Higgs
field candidates and massless 4d fermions are generated at the tree level,
and this appears to be a rather general property of the $Z_2$-invariant backgrounds in models with simple
underlying gauge groups. It can be evaded if the background explicitly breaks the $Z_2$-symmetry.

The second example involves extra $U(1)$ factor in the gauge group
of the multi-dimensional theory and leads to a toy model of leptons.
Three left-handed and one right-handed
generations are obtained from a single fermion of the underlying theory,
a property analogous to the multiplication of fermion generations in
models with fermion localization on topological defects with non-minimal
topological numbers~\cite{ml}. To obtain two more right-handed ``leptons''
one adds extra fermions into multi-dimensional theory. In this example,
only one fermion obtains a mass upon electroweak symmetry breaking,
which again illustrates how constrained is the structure of
the Yukawa sector.

Our examples have only illustrative nature and by no means pretend to be close to realistic extensions of
the Standard Model, although they contain
 the correct spectrum of the standard electroweak theory of leptons.
 It remains to be understood whether our construction
can be used for successful model-building.


\section{Higgs from gauge fields}

In this paper we consider 
8-dimensional space-time, although the 
scheme  can be
generalized to 10 and higher dimensions.
Let us consider  the Yang--Mills theory with the action
\[
S= - \frac{1}{4g^2} \int d^8 X ~\sqrt{-G}~  \mbox{Tr} ( F_{MN}F^{MN}) \; ,
\]
where $G$ is the determinant of the 8-dimensional metric.
The class of models 
of interest to us has the 8-dimensional
space-time of the product form  $R^4 \times M_2\times M' _2$,
where $M_2$ and $M_2^\prime$ are 2-dimensional 
compact manifolds.
The local coordinates on $M_2$ and $M^\prime_2$ are denoted by $y^m$,
$m=1,2$ and
 $y^{ m^\prime}$, $m^\prime=1^\prime,2^\prime$.
We will often use  complex combinations
$z= y^1+iy^2, \bar z = y^1-iy^2$ and similarly for  $z', \bar z' $.
By an appropriate choice of coordinates the metrics on $M_2$ and
$M' _2$ can, locally,   be brought to the Gaussian form,
\beq
ds^2 = \psi^2 (z, \bar z ) dz d\bar z, \quad\quad\quad\quad ds{'^2}
= \psi{'^2} (z', \bar z' ) dz' d\bar z',
\label{metric}
\eeq
We will treat $M_2$ and $M'_2$ symmetrically: whatever we
say about $M_2$ will also be valid for $M'_2$.
Henceforth we  concentrate on $M_2$.

The background configuration
\beq
\bar F_{mn} = \frac{2\pi n}{\Omega_2} \, \psi^2 \varepsilon_{mn} X
\label{Fbar}
\eeq
 solves the Yang--Mills equations on $M_2$. Here   $\varepsilon_{12} =+1$,
 $n$ is a constant, $\Omega_2$ is the volume of $M_2$ and $X$
is a generator of the gauge group.
With appropriate normalization of $X$, $n$ takes integer values,
so that the flux is quantized.
We take the background gauge field on $M'_2$ to be in the same
direction $X$ in
the Lie algebra; this field is characterized by an integer $n^\prime$.
All  other components of the background gauge field
are set equal
to zero.

The bilinear part of the action for perturbations becomes
\[
S= - \frac{1}{2g^2} \int d^8 X ~\sqrt{-G}~
\mbox{Tr} \{ D_M V_N D^M V^N - D_M V_N D^N V^M - i\bar F_{MN} [V_M,V_N]\}
\]
where  $V_M$ are the gauge field fluctuations, so that
 $ A_M= \bar A_M + V_M$, and the covariant
derivative is given by
\[
D_M V^N =  \nabla _M V^N - i[\bar A_M, V^N]
\]
with
$\nabla_M $ being
the standard Riemannian covariant derivative.
We are interested in the components $V_m$ and $V_{m'}$
which are tangent to $M_2$ and $M^\prime_2$, respectively. These are
  vector fields
on $M_2$ and $M_2^\prime$, respectively,  while from the standpoint of
$R^4$ they make   KK towers of scalar fields.
If some KK modes of $V_m$  and/or $V_{m'}$
are light, they become the Higgs field candidates.
Let us see under which conditions this indeed happens.

For the sake of argument,
let us assume that $V_m$ and $V_{m^\prime}$ decouple from
each other; this assumption will be justified {\it a posteriori}.
Then it suffices to consider the fields $V_m$ only.
It is convenient to use
 an orthonormal frame in the tangent space of $M_2$.
We denote the indices in this frame by $a, b,...=
\underline 1, \underline 2$;
underlining here signifies that the indices refer
to the orthonormal frame.
For the metric in the form (\ref{metric}), this simply means that
$V_{\underline 1} = \psi^{-1} V_1$,
$F_{\underline 1 \underline 2} = \psi^{-2} F_{12}$, etc.
Note that in this notation, the backgorund fields $\bar{F}_{ab}$ and
$\bar{F}_{a^\prime b^\prime}$ are constants on $M_2$ and $M_2^\prime$,
see Eq.~(\ref{Fbar}).

The explicit form of the bilinear action for $V_a$ is
 \beq
S= -\frac{1}{2g^2} \int d^8 X \sqrt{-G}  \mbox{Tr}
\{ \partial_\mu V_a \partial^\mu V^a  + D_bV_a D^b V^a
+ D_{b' }V_a D^{b' }V^a -D_bV_a D^a V^b   - i\bar F_{ab} [V_a,V_b] \}
\label{actionlong}
\eeq
In the complex basis in the tangent space one has
$ V_- = V_{\underline 1} - iV_{\underline 2} = \psi^{-1} V_{z}$, $V_+
= (V_{-})^{\dagger} $.
Likewise $D_{-}=  D_{\underline 1} - i D_{\underline 2}$.
For quantities in the complex basis of $M_2^\prime$ we will use the
notation $V^\prime_-$, $D^\prime_-$, etc.; as an example,
 $D^\prime_- =  D_{\underline 1^\prime} - i D_{\underline 2^\prime}$.

By inspection of the expression (\ref{actionlong}) one observes
that potentially light modes satisfy
\beq
D_+ V_- = D'_{+} V_- = 0 \; .
\label{maineq}
\eeq
For such fields the bilinear action simplifies to
 \beq
S= - \frac{1}{2g^2} \int d^8 X \sqrt{-G}
\mbox{Tr} \{ \partial_\mu V_+ \partial^\mu V_-
+ i V_+ [\bar F'_{+-} - \bar F_{+-},V_-] \}
\label{keyaction}
\eeq
 The key point is that the masses of the complex fields  $V_\pm$
are given by the difference $(\bar F'_{+-} - \bar F_{+-})$.
This is precisely the cancellation found in Ref.~\cite{Seif-Gia}:
once the background is chosen in such a way that
this difference is small, the fields obeying (\ref{maineq}) are light,
even though each of $\bar F_{+-}$ and $\bar{F}_{+-}^\prime$ is large.
Mass terms for all other KK modes of $V_m$
receive contributions from the Laplacian acting on  $V_\pm$ and therefore for
small size $M_2$ and $M'_2$ these other modes will have large masses.
In this way a finite number of
light scalars on $R^4$ --- the Higgs field candidates ---
will be separated from an infinite number
of heavy KK modes. 

Before discussing solutions to Eq.~(\ref{maineq}), let us make a few remarks.

1. The fields obeying  Eq.~(\ref{maineq}) satisfy $D_a V^a =0$. Due to this property,
the light modes of $V_m$ indeed decouple from the perturbations $V_{m^\prime}$.
This justifies the assumption we made above.

2.  As we already pointed out, for the background field configuration given in
Eq.~(\ref{Fbar}), $F_{+-}$ and $F'_{+-}$ are constants on $M_2$ and $M_2^\prime$.
Therefore, if Eq.~(\ref{maineq}) has several solutions of one and the same $X$-charge,
masses of all these light
modes are equal to each other.
Furthermore, if equations
\beq
D'_+ V'_- = D_{+} V'_- = 0 \; .
\label{maineq-prime}
\eeq
also have  non-trivial solutions of the same $X$-charge,
the light modes of $V$ and $V'$ have masses squared
equal in absolute value
but opposite in sign: if one of them is tachyonic,
another is necessarily normal, and {\it vice versa}.

3. If 
$     \bar{F}^\prime_{+-} = \bar{F}_{+-} $,
the Higgs field candidates are massless at the tree level. 
Equality between $\bar{F}_{+-}^\prime$ and $\bar{F}_{+-}$ may
be either due to fine tuning, or a consequence of a $Z_2$-symmetry interchanging
the two manifolds $M_2$ and $M^\prime_2$ and the background fields on them.
In the latter case one has
\[
   n = n^\prime \;, \;\;\;\;\; \Omega_2 = \Omega_2^\prime \; .
\]
We think this $Z_2$-protection
 is a particularly interesting property of the class of models discussed
in this paper. Clearly, in the case of $Z_2$-symmetry, both $V_-$ and $V^\prime_-$ have
massless modes 
at the classical level. If this symmetry is slightly broken at the classical level,
so that  $\bar{F}^\prime_{+-} \neq \bar{F}_{+-}$,
the tree level masses of both $V_-$ and $V^\prime_-$ are small; the number of light scalars
generated in this way is larger than 1. In fact, we will see
in what follows that this
$Z_2$-protection  mechanism naturally gives rise to fairly large number of light
scalar modes (either massless, or normal and tachyonic in equal number).
Alternatively, one can think of generating the masses for the Higgs candidates
radiatively. 
We will further discuss this issue at the end of this paper.

4. Like in other gauge-Higgs unification 
models~\cite{Seif-Gia,GHU3,Csaki1,Scrucca1}, and
unlike in the case of the Hosotani/Scherk--Schwarz mechanism,
quartic interactions of
the Higgs candidates are present at the tree level. The reason is that
solutions  of Eq.~(\ref{maineq}) have both $V_1$ and $V_2$ non-zero,
and their commutator does not vanish. Thus, the quartic Higgs self-coupling
$\lambda$ is generically of order $\lambda \sim g_4^2$, where $g_4$ is the
4-dimensional 
gauge coupling. 
This is not only relevant to phenomenology, but also ensures
the self-consistency of the entire approach. Indeed, there are no topological 
arguments guaranteeing the stability of the background (\ref{Fbar}), so 
whether or not the 8-dimensional field
configutrations we discuss are stable is a dynamical issue.
For low masses of the candidate Higgs bosons, $m_H^2 \ll 1/R^2$,
perturbative stability can be analysed within 4-dimensional
low energy theory (KK modes have positive masses squared).
Since the Higgs self-coupling does not vanish, the Higgs vacuum ---
and hence the entire 8-dimensional solution --- is perturbatively stable.
It is clear, though, that we are dealing with a metastable
vacuum. Its decay rate, however, is expected to be suppressed as
$\mbox{exp} (- \mbox{const}/g_4^2)$, thus making our models potentially 
viable.

Let us now proceed to solving  Eq.~(\ref{maineq}), still in rather general
terms. The field $V_-$ is a linear combination of the generators of the gauge
algebra with complex (space-time dependent) coefficients. So, one can view
this field as belonging to complexified adjoint representation of the gauge algebra.
In this representation one can choose the basis $T^i$ in such a way that
$[X, T^i]= q^i T^i$, where $q^i$ are real, and decompose $V_-$
as  $ V_- = V_- ^i T^i$. Obviously, the meaning of $q^i$
is that it is equal to the $X$-charge of the corresponding Higgs candidate.
We will omit the superscript $i$ in what follows.

For given $X$-charge $q$,  the first and the second of Eq.~(\ref{maineq}) read
 \beqar
 \partial _{\bar z} V_-   + (\partial _{\bar z}\, \ln \psi) \, V_-  -i q \bar{A}_{\bar z}
(z, \bar z) V_- &=&0
 \nonumber \\
 \partial _{\bar z'} V_-    -i q \bar{A}_{\bar{z}'}(z', \bar z') V_- &=&0
\nonumber
\eeqar
These  first order equations  are solved by 
 \beq
V_- = \frac{1}{\psi (z, \bar z')} \,
\mbox{exp}\left\{ iq \int  d\bar z \bar{A}_{\bar z} (z,\bar z)\right\} \,
\mbox{exp}\left\{ iq \int  d\bar z' \bar{A}_{\bar{z}'} (z',\bar z')\right\} \, f (z) \, g (z')
\label{15}
 \eeq
 where $f$ and $g$ are holomorphic functions of their arguments.  These functions
should be chosen in such a way that $V_- $ is normalizable, namely,
 \[
 \int d^2 z \psi^2  V_+ V_- \;\;\; ,\quad\quad\quad\quad
 \int d^2 z'  \psi'^2 V_+ V_-
 \]
 are both finite. These conditions  restrict the number of permissible solutions.

To proceed further, let us assume that $M_2$ and $M_2^\prime$ are Einstein manifolds,
i.e., $R_{mn} = \mbox{const} \cdot g_{mn}$.
Then the solutions  to the Yang--Mills equations on $M_2$ and $M_2'$,
with the field strength  given by (\ref{Fbar}), are
\[
A_m = \frac{n}{2}  \omega_m X \; , \;\;\;\; A_{m'}= \frac{n'}{2} \omega_{m'} X
\]
where $\omega _m = \varepsilon_{ml} \partial_l \ln \psi$ and
 $\omega_m'=\varepsilon_{m' l'} \partial_{l'} \ln \psi'$  are the
components of spin connections in $M_2$ and $M'_2$,  respectively.
The solutions (\ref{15}) then take simple form,
\[
   V_- = \psi^{\frac{qn}{2}-1} (z, \bar z) \, \psi^{\prime \,\, \frac{qn'}{2}} f(z) g(z') \;.
\]
The norms on $M_2$ and $M_2^\prime$  reduce to
 \beqar
  \int d^2 z \psi^2  V_+ V_- &=& \int d^2 z \psi^{qn} |f|^2
\label{normM} \\
  \int d^2 z' \psi'^2  V_+ V_- &=& \int d^2 z' \psi^{\prime \, 2+qn'} |g|^2
\label{normMprime}
\eeqar
Clearly, the number of solutions of finite norms is finite. As we will
see momentarily, this number can be easily counted by making use of the
above formulas. We note in passing that there are no normalizable solutions of zero $X$-charge
$q$.

The same analysis, with obvious interchange $z \leftrightarrow z'$,
etc., applies to the light modes of $V^\prime_m$.

 \section{Examples}
To illustrate the general treatment, let us give  simple examples
of  models 
leading to the light scalars with quantum numbers of the
Higgs field of the Standard Model. We also introduce fermions in such a
way as to mimic leptons.

\subsection{$SU(4)$}

We begin with the gauge group $SU(4)$
and $X = \mbox{diag} (1,1,0,-2)$.
The background breaks
$SU(4)$ down to $SU(2)_L\times U(1)_X\times U(1)$ in such a way that
the complexified adjoint of $SU(4)$ is decomposed as
 \[
 \underline {15} =   \underline{3}_0+1_0 +  \underline{1}_0 +
 \underline{2}_{3 } +  \underline{2}_{-3} +  \underline{2}_{1} +  \underline{2}_{-1}  +
 \underline{1}_{2} + \underline{1}_{-2}
 \]
 the subscript here refers to the $X$-charge.
 Clearly $ \underline{2}_{1}$ is the right candidate for the Higgs doublet,
once $X$ is identified with the weak hypercharge\footnote{ In
our normalization
the  left-handed lepton doublet has  weak hypercharge  -1 and the right-handed
electron has weak hypercharge  -2.}, $Y=X$.

Let us take $M_2=S^2$ and $M'_2=S'^2$ with the radii $a$ and $a'$ respectively.
The metric functions are
\[
\psi = \frac{1}{1+ \frac {|z|^2}{4a^2}} \;\;\; ,
\quad\quad\quad \psi'= \frac{1}{1+ \frac {|z'|^2}{4a'^2}}
\]
It is now straightforward to count the number of light scalars in this setup.
Let us take $n>0$ and $n'>0$. Then,
according to Eqs.~(\ref{normM}) and (\ref{normMprime}), all light scalars
must have positive hypercharges. Let us
specify to doublets of hypercharge $q=1$. For convergence of the integral
(\ref{normM}), $n$ should be larger than $1$.
The choice  $f= z^m$ and $g=z'^{m'}$
yields convergent norms provided that $m= 0,1,\dots (n-2)$,
and $m'= 0, 1,\dots n' $.
Thus there are $(n-1)(n'+1)$ normailzable solutions in the sector of the gauge
fields tangent to $M_2$. Since the background is invariant under the rotations of the two spheres, 
the entire spectrum and the interactions are
 classified according to irreducible 
representations\footnote{The relationship between the 
holomorphic basis used in this paper and  spherical harmonic basis of Ref.~\cite{Seif-Gia}  
can be found in
Ref.~\cite{Karabali:2004xq}.} of 
$SO(3)\times SO(3)'$ acting on $S^2\times S'^2$.
The light Higgs belongs to $(j= -1+ n/2 ,  j'= n'/2)$ representation of
this group. Likewise, for $n'>1$ there is $(n'-1)(n+1)$
light modes among the gauge fields tangent to $M_2^\prime$ belonging
 to  $(j=n/2 ,  j'=-1+n'/2)$ representation of $SO(3)\times SO(3)'$.

Altogether,
there are $2(nn'-1)$ Higgs candidates. When 
the  $Z_2$-symmetry 
is slightly broken at the classical level,
half of these putative Higgs fields
will be tachyonic and the other
half will have positive mass squared, where the absolute value of
$mass^2$  is given by $|n/a^2 -n'/a'^2|$.

In the case of $Z_2$-symmetric setup, when $n=n'$, the number of the light
scalars
is either zero (for $n,n' = 0,1$) or at least 6 (for $n=n'=2$).
This illustrates the fact that the $Z_2$-protection mechanism leads to
rather large number of light 
scalar fields descending from the multi-dimensional
gauge field.

Let us now consider fermion zero modes in this example.
We start from an 8-dimensional spinor in the
anti-fundamental representation of $SU(4)$, which
after symmetry breaking by the background field is decomposed as
$\underline{\bar{4}} = \underline{2}_{-1} + \underline{1}_2 + \underline{1}_0$.
A singlet of hypercharge 2 and a doublet have quantum numbers of left-handed positron and left-handed
leptons of the Standard Model. Let us see that there are corresponding zero modes.

The 8-dimensional Dirac equation reads
 \[
 \Gamma^A E_A ^M \left( \partial_M + \frac{1}{2} \omega_{M[ CD]}
\Sigma^{[CD]} -i \bar{A}_M\right)\chi=0
 \]
 where $\Gamma^A$ are $16\times 16$ constant Dirac matrices,
$  \Sigma^{[CD]}= \frac{1}{4} [\Gamma^A, \Gamma^B]$ are the generators of $SO(1,7)$ and
$\omega_{M[CD]}$ are the components of the spin connection.  Let us take
the 8-dimensional
spinor $\chi$ to be chiral,
  \beq
   i\Gamma_0\Gamma_1....\Gamma_7 \chi = +\chi
\label{chichi}
\eeq
The zero mode equations on $M_2$ and $M'_2$ reduce to
   \beqar
   (\partial_4 +\Gamma_{\underline 4\underline 5} \partial_5)\chi
+\frac{1}{2}
[(\partial_4 +\Gamma_{\underline 4\underline 5} \partial_5) \ln \psi]
\, (1+iqn\Gamma_{\underline 4\underline 5})\chi &=&0
\label{zero-onM} \\
   (\partial_6 +\Gamma_{\underline 6\underline 7} \partial_7)\chi
+\frac{1}{2}
[(\partial_6 +i\Gamma_{\underline 6\underline 7} \partial_7) \ln \psi']
\, (1+iqn'\Gamma_{\underline 6\underline 7})\chi &=&0
\label{zero-onMprime}
\eeqar
where $q$ is now the fermion hypercharge.
Let $\epsilon=\pm 1$ and $\epsilon'=\pm 1$ be eigenvalues of $i\Gamma_{\underline 4\underline 5}$ and
$i\Gamma_{\underline 6\underline 7}$, that is, chiralities on $M_2$ and $M_2'$,
respectively. Then the solutions to Eqs.~(\ref{zero-onM}) and (\ref{zero-onMprime})
are
   \beq
   \chi =  \left( 1+\frac{|z|^2}{4a^2}\right)^{\frac{1+\epsilon qn}{2}}
\left( 1+\frac{|z'|^2}{4a'^2}\right)^{\frac{1+\epsilon' qn'}{2}} z^m_\epsilon z'^{m'}_{\epsilon'}
\chi^{m,m'}_{\epsilon \epsilon'}
\label{resultchi}
   \eeq
where $z_\epsilon = \bar{z}$ for $\epsilon=+1$ and $z_\epsilon = z$ for $\epsilon=-1$,
while $\chi_{\epsilon \epsilon'}$ is a spinor 
independent of $z, \bar z , z'$ and $\bar z'$ and satisfying the 4-dimensional
chiral Dirac equation. It follows from (\ref{chichi})
that its 4-dimensional chirality is  $(-\epsilon \epsilon')$.
We see from (\ref{resultchi})
that zero modes exist for $q\neq 0$ only, and that their chiralities
$\epsilon$ and $\epsilon'$ 
must be both negative for $q>0$ and both positive for $q<0$.
In either case,  the 4-dimensional chirality
is negative. Fermions of positive 8-dimensional chirality
have zero modes which are left-handed from 4-dimensional viewpoint.

The zero modes have to be normalizable, that is the following integrals have to be finite,
\[
\int~dz d\bar{z} \psi^2 \bar{\chi} \chi \; , \;\;\;\; \int~dz'd\bar{z}' \psi^{\prime 2}
\bar{\chi} \chi \; .
\]
This restricts the number of zero modes. In particular, there are $n n'$ zero modes of
hypercharge $\pm 1$ and $4 n n'$ zero modes of hypercharge $\pm 2$. We see that models considered in
this paper allow for several fermionic generations originating from single multi-dimensional
fermion.

The major problem with this model is that all 4-dimensional fermions
have the same 4-dimensional chirality. Since the original gauge interactions do not
involve charge-conjugate fermions, this means that Yukawa interactions between zero fermion
modes and the light scalars --- extra-dimensional components of the gauge field --- are absent
at least at the tree level. More generally, fermions of the same sign of hypercharge have
the same 4-dimensional chiralities, which is not the case in the Standard Model.
To construct a model with non-zero Yukawa couplings one has to cure this problem.

\subsection{$U(3)\times U(1)$}

To obtain fermions of both positive and negative 4-dimensional
 chiralities and non-vanishing
Yukawa couplings in a $Z_2$-symmetric background, we modify the previous example by adding a $U(1)$ factor to the
gauge group. To avoid minor but unnecessary complications, it is convenient to
consider $U(3)$ instead of $SU(4)$ of the previous example. Thus, in our second
 example the gauge group is $U(3) \times U(1)_{\tilde{X}}$, and
$X = \mbox{diag} (1,1,0) \in U(3)$.

The treatment of light scalar doublets is the same as in the
previous example, so we concentrate on fermions.
We choose them to have positive 8-dimensional chirality as in (\ref{chichi}) and begin with
$(\underline{3}, -1)$ representation of $U(3) \times U(1)_{\tilde{X}}$. Upon
symmetry breaking by the background, $U(3)\times U(1)_{\tilde{X}} \to
U(2) \times U(1)_X \times U(1)_{\tilde{X}}$,  it decomposes as
\beq
  (\underline{3}, -1) = (\underline{2}, 1_X, -1_{\tilde{X}}) + (\underline{1}, 0_X,  -1_{\tilde{X}})
\; .
\label{odot}
\eeq
If one wishes to identify these fermions with left-handed lepton doublets and right-handed
lepton singlets of the Standard Model, one makes the assignment of weak hypercharge $Y= X + 2\tilde{X}$.
With this assignment, the Higgs doublets still have weak hypercharge $Y=1$.
Note, however, that unlike in the Standard Model, the low energy gauge group
is $U(2) \times U(1)_Y \times U(1)_{\tilde{Y}}$, where $\tilde{Y}$ is the second
linear combination of $X$ and $\tilde{X}$.

Now, the trick is to populate the internal manifolds $M_2$ and $M_2^\prime$
with both $U(1)_X$ and $U(1)_{\tilde{X}}$ gauge fields. This can still be done in
a $Z_2$-symmetric way, with $M_2 \leftrightarrow M_2^\prime$,  $F \leftrightarrow F^\prime$
and $\tilde{F} \leftrightarrow - \tilde{F}^\prime$ under the $Z_2$-transformation.
Due to this symmetry, the topological numbers $n$,$n'$ of $U(1)_X$ and $\tilde{n}$,
$\tilde{n}'$ of $U(1)_{\tilde{X}}$ are related as $n'=n$, $\tilde{n}' = - \tilde{n}$.
The fermion doublet effectively feels Abelian fields on $M_2$ and $M_2^\prime$ with
\[
   (qn)_D = n - \tilde{n} \; , \;\;\;\; (qn^\prime)_D = n+ \tilde{n} \; ,
\]
while for the singlet one has
\[
   (qn)_S = - \tilde{n} \; , \;\;\;\; (qn^\prime)_S = \tilde{n} \; .
\]
Clearly, $(qn)_S$ and    $(qn^\prime)_S$ have opposite signs, while the signs of
 $(qn)_D$ and    $(qn^\prime)_D$ can be made the same by an appropriate choice of $n$ and
$\tilde{n}$. 
By making use of
 Eq.~(\ref{resultchi}) one finds that
in that case the singlet zero modes are right-handed while the doublet ones are left handed.

Let us further specify to the simplest case
\beq
    n = 2\; , \;\;\;\;\; \tilde{n} = 1
\label{star}
\eeq
Then there are 3 left handed zero modes $D^{m,m'}$, with 
$m=0$, $m'=0,1,2$ and $\epsilon_D
= \epsilon_D^\prime = -1$, where $m$ and $m^\prime$  are the integers entering
(\ref{resultchi}). The singlet has one right-handed zero mode $S^{0,0}$ with
$\epsilon_S=+1$, $\epsilon_S^\prime = -1$,
and the Higgs
candidates are  $V_-^{0,\,m^\prime}$, $m'=0,1,2$ and $V_-^{\prime \, m,\, 0}$,
$m=0,1,2$.  The $SO(3)\times SO(3)'$ quantum numbers of these modes are respectively
$D$:~$(j=0, j' =1)$, $S$:~$(j=0, j'=0)$, $V$: $(j=0, j'=1)$ and $V'$: $(j=1, j'=0)$. Upon integration
over $S^2\times S'^2$ the term
\[
  \bar{D} V_- (\Gamma_4 + i \Gamma_5)S
\]
produces a non-zero Yukawa coupling, while the Yukawa coupling with $V^\prime_{\pm}$ is
forbidden by $SO(3) \times SO(3)^\prime$ symmetry.
As the Yukawa terms arise from the interactions of fermions with $U(3)$ gauge fields in the
8-dimensional theory, 
the Yukawa couplings are  of order of the 4-dimensional
gauge coupling. In fact, 
after  integrating over $S^2\times S'^2$ and rescaling 
to canonically normalized fields one finds that 
the Yukawa coupling is equal to $2g$ 
where $g$ is the 4-dimensional $U(2)$-coupling.  
Note that when the Higgs fields get vacuum expectation values, 
only one
fermion obtains a mass, simply because 
there is only one right-handed fermion coming from
(\ref{odot}).

One way to obtain a toy model of leptons is to add two fermionic $U(3)$-singlets
with $\tilde{X} = -1$, again of positive 8-dimensional chirality. These singlets will
form two $(\underline{1}, 0_X, -1_{\tilde{X}})$ representations and will have one
right-handed zero mode each. Since they do not interact with $U(3)$ gauge fields, their
Yukawa couplings will be zero at the tree level. In this way one obtains three families
with lepton quantum numbers, only one of them having a tree-level mass after
electroweak symmetry breaking.

Clearly, the model discussed here is far from being close to realistic.
It does not contain quarks, its low energy gauge group is $U(2) \times U(1) \times U(1)$,
it has global $SO(3) \times SO(3)$ symmetry\footnote{This symmetry
will be promoted to gauge symmetry when gravitational interactions are
included. It would 
of course be absent if $M_2$ and $M'_2$ had no isometry groups.} 
inherited from the internal manifold
$S^2 \times S^{\prime \, 2}$, etc. Nevertheless, this model illustrates
that our construction has some features which we think are interesting.

Except for special features mentioned earlier, phenomenology of the
class of models we discuss appears rather similar to other models of
gauge--Higgs unification in more than 5 dimensions. Naive dimensional
analysis, extended to higher dimensions~\cite{Chacko:1999hg}, suggests
that the UV cutoff scale $\Lambda$
of the 8-dimensional theory is determined by
$\Lambda^4 \simeq l_8 g^{-2}$, where $g$ is the 8-dimensional gauge
coupling and $l_8 = 3! \, 2^8 \, \pi^4$ is the  8-dimensional
loop factor. In terms of the size of extra dimensions
$a$ one has
\[
\Lambda \simeq \frac{l_8^{1/4}}{\sqrt{4\pi g_4} a} \sim \frac{10}{a} \; .
\]
As we already pointed out, the candidate Higgs bosons may obtain
tree-level masses (normal and tachyonic), if the $Z_2$-symmetry is
broken explicitly by the background. Otherwise these masses may come
from higher-order operators like $l_8^{-1} \mbox{Tr} F^4$ and from 
non-local operators induced radiatively; neither of these types of
contributions to the masses
is forbidden by the $Z_2$-symmetry. The former contributions
are estimated as $(m_H^2)_{H O} \simeq (2\pi^2 g_4^2)/(l_8 a^2)$ (for
$n, n^\prime \sim 1$), while the latter are expected to
be somewhat larger, 
$(m_H^2)_{N  L} \simeq g_4^2/(l_4 a^2)$ where $l_4 = 16 \pi^2$
is the 4-dimensional loop factor. Thus, our construction belongs to
the class of theories with TeV-scale extra dimensions 
fairly low cutoff scale
 $\Lambda$, about 10~TeV or somewhat higher.
It remains to be explored how far one can go in model-building with this
construction.

\section*{Acknowledgements}

The authors are indebted to D.~Gorbunov, V.P.~Nair,
E.~Nugaev and S.~Troitsky for useful discussions.
The work of V.R. is supported in part by the Russian Foundation 
for Basic Research grant No. 05-02-17363a.

\end{document}